\pdfoutput=1

\documentclass[11pt]{article}

\usepackage[preprint]{acl}

\usepackage{times}
\usepackage{latexsym}

\usepackage[T1]{fontenc}

\usepackage[utf8]{inputenc}

\usepackage{microtype}

\usepackage{inconsolata}

\usepackage{graphicx}
\usepackage{tcolorbox}
\usepackage[
    colorinlistoftodos,
    textsize=footnotesize,
        ]{todonotes}

\usepackage[
    colorinlistoftodos,
    textsize=footnotesize,
        ]{todonotes}
\definecolor{myblue}{RGB}{10, 70, 180}  

\usepackage{subcaption} 
\usepackage{tcolorbox}
\tcbuselibrary{listingsutf8}
\usepackage{amsmath}
\usepackage{amsfonts}
\usepackage[utf8]{inputenc}

\usepackage{listings}
\usepackage{booktabs}

\title{Noise, Adaptation, and Strategy: \\Assessing LLM Fidelity in Decision-Making}

\author{
  Yuanjun Feng\textsuperscript{1} \quad
  Vivek Choudhary\textsuperscript{2} \quad
  Yash Raj Shrestha\textsuperscript{1} \\
  \textsuperscript{1}University of Lausanne \\
  \textsuperscript{2}Nanyang Technological University \\
  \texttt{\{yuanjun.feng, yashraj.shrestha\}@unil.ch} \\
  \texttt{vivek.choudhary@ntu.edu.sg}
}

\begin{document}
\maketitle
\begin{abstract}
Large language models (LLMs) are increasingly used in social science simulations. While their performance on reasoning and optimization tasks has been extensively evaluated, less attention has been paid to their ability to simulate human decision-making's variability and adaptability. We propose a process-oriented evaluation framework with progressive interventions (\emph{Intrinsicality}, \emph{Instruction}, and \emph{Imitation}), to examine how LLM agents adapt under different levels of external guidance and human-derived noise. We validate the framework on two classic economics tasks, irrationality in the second-price auction and decision bias in the newsvendor problem, showing behavioral gaps between LLMs and humans.

We find that LLMs, by default, converge on stable and conservative strategies that diverge from observed human behaviors. Risk-framed instructions impact LLM behavior predictably but do not replicate human-like diversity. Incorporating human data through in-context learning narrows the gap but fails to reach human subjects' strategic variability. These results highlight a persistent alignment gap in behavioral fidelity and suggest that future LLM evaluations should consider more process-level realism. We present a process-oriented approach for assessing LLMs in dynamic decision-making tasks, offering guidance for their application in synthetic data for social science research.
\end{abstract}

\section{Introduction}
Large language models (LLMs) are increasingly applied to tasks requiring decision-making, planning, and reasoning \cite{rosenman_llm_2024, choi_people_2025, huang_can_2024}. As interest grows in using LLMs to simulate human subjects in social science, recent work has moved beyond static tasks toward more dynamic and interactive evaluations \cite{gueta_can_2025}. Benchmarks like MoralBench \cite{ji_moralbench_2024} and the Decision-Making Behavior Evaluation Framework \cite{jia_decision-making_2024} assess LLMs on single-shot tasks such as ethical dilemmas, risk preferences, and loss aversion. Similarly, economic game studies (e.g., Dictator, Ultimatum, Public Goods) show that LLMs can reproduce some human-like behaviors, such as generosity or cooperation \cite{akata_playing_2025, mozikov_eai_2024}. These evaluations typically focus on final choices or performance. However, human decisions are often noisy, history-dependent, and shaped by bounded social and cognitive constraints \cite{santos_evolutionary_2015}. While LLMs now match or surpass human accuracy on standard reasoning benchmarks \cite{peng_survey_2024, leng_llm_2024}, their ability to reproduce these stochastic patterns remains an open question:

\begin{tcolorbox}[colframe=black, colback=white, boxrule=0.2mm]
\textbf{To what extent do LLMs exhibit behavior consistent with human decision-making, and can this behavior be modulated through targeted interventions?}
\end{tcolorbox}

As LLMs are increasingly proposed for use in behavioral modeling, synthetic data generation, and experimental simulation, it is crucial to assess whether their behavior reflects these foundational properties of human cognition \cite{wang_aligning_2023}. Our study contributes to this goal by introducing a structured framework for evaluating behavioral alignment between LLMs and humans in dynamic decision-making contexts.

\begin{figure*}[htbp]
  \centering
  \includegraphics[width=0.9\textwidth]{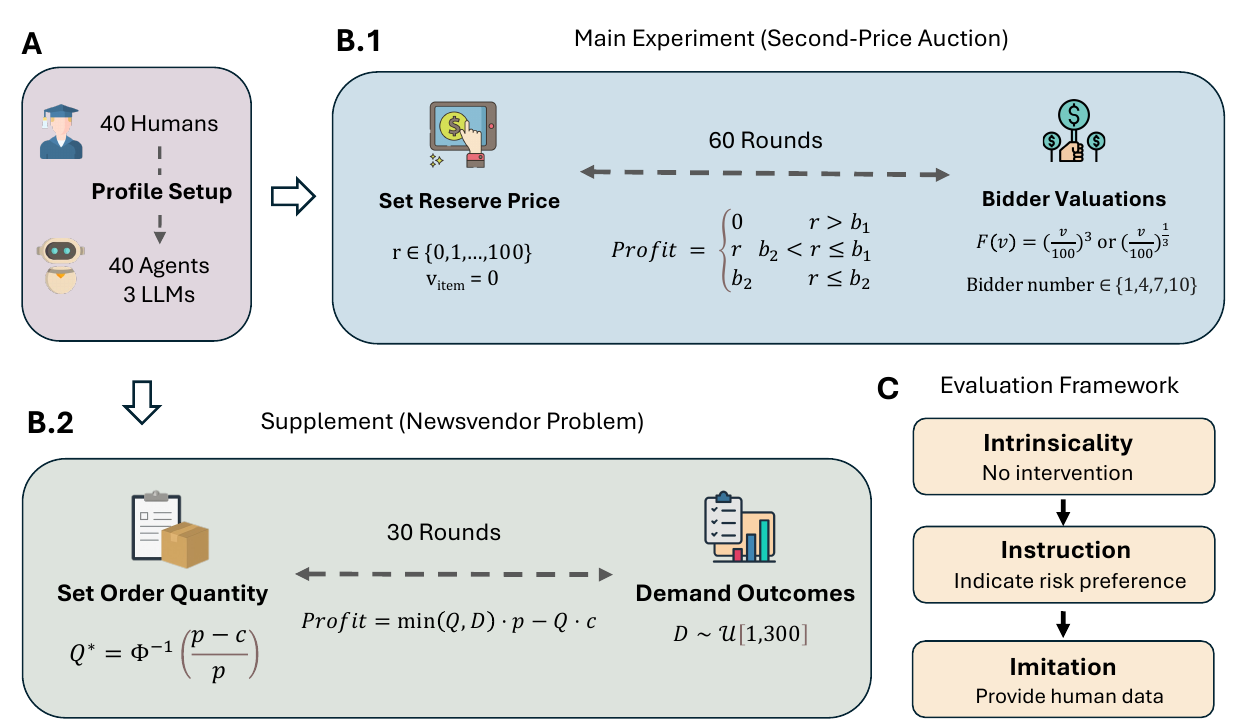}
  \small
  \caption{
  Overview of the experimental design. (\textbf{A}) We instantiate 40 agents per LLM with real human demographic profiles. (\textbf{B.1}) In the main experiment (second-price auction), agents set a reserve price $r$ in 60 rounds and receive simulated bidder valuations drawn from known distributions. Profit depends on bidder valuations $b_1$ and $b_2$ (highest and second-highest bids). (\textbf{B.2}) In the supplementary experiment (newsvendor problem), agents choose an order quantity $Q$ over 30 rounds to maximize expected profit under stochastic demand. $p$ is the selling price, and $c$ is the cost. (\textbf{C}) On both tasks, we apply an evaluation framework with progressive interventions: \textbf{Intrinsicality} (no intervention), \textbf{Instruction} (indicate risk-preference), and \textbf{Imitation} (provide historical human data).}
  \label{fig:study_flow}
\end{figure*}


To systematically evaluate LLM behavior, we propose a process-oriented evaluation framework with progressive interventions: (1) \textbf{Intrinsicality}, LLMs operate without any intervention; (2) \textbf{Instruction}, LLMs receive risk-framed instructions; and (3) \textbf{Imitation}, LLMs receive partial human decision histories and are tasked with continuing the behavior. This framework enables us to assess the extent to which LLMs exhibit key features of human decision-making, such as bounded rationality (taking suboptimal strategies under limited cognitive resources) or behavioral variance (individual variability in decisions, often linked to risk preference or adaptation).

We apply this framework to two classic behavioral tasks: a second-price auction \cite{edelman_internet_2007, cooper_understanding_2008}, where subjects set reserve prices before observing bids, and a newsvendor problem \cite{schweitzer_decision_2000}, where subjects choose order quantities for newspapers under uncertain demand. Both tasks feature dynamic feedback and closed-form optimal strategies, enabling direct comparison between LLM and human decision patterns. We instantiate LLM agents using GPT-4o, Claude 3.5 Sonnet, and Claude 3.7 Sonnet, and compare their behaviors with that of human subjects.

Since the second-price auction has established theoretical benchmarks and involves bilateral interaction and strategic complexity, we designate it as the primary use case. We conduct a comprehensive evaluation of LLM behavior on this task, comparing it against empirical human data and documented behavioral theories. The newsvendor task serves as a supplementary experiment to verify the generalizability of our framework. We illustrate the two experimental processes and the evaluation framework in Figure~\ref{fig:study_flow}.

\textbf{Contributions:} We introduce a process-oriented evaluation framework with progressive interventions to systematically assess whether LLM agents exhibit the stochasticity and adaptiveness characteristic of human decision-making. Through two classic behavioral experiments (second-price auction and the newsvendor problem), we demonstrate that LLMs consistently display low-variance, highly stable strategies, with minimal within-agent fluctuation or cross-agent diversity. These findings highlight fundamental limitations in using current LLMs as synthetic proxies for human subjects in dynamic behavioral settings and provide a practical framework for auditing LLM behavior in decision-making tasks.

\section{Background and Related Work}

\subsection{LLMs as Subjects in Social Science}

LLMs are increasingly employed as proxies for human subjects in experimental research across domains such as psychology \cite{binz_using_2023}, political science \cite{liu_turning_2025}, and behavioral economics \cite{ross_llm_2024}. Researchers often apply LLMs to tasks involving moral reasoning, social forecasting, and decision-making \cite{bankins_multilevel_2024}, where models frequently perform at levels comparable to humans. For example, \citet{chiang_can_2023} show that LLMs match human experts in evaluation and reasoning during open-ended story generation and adversarial attacks. In a large-scale replication study involving 154 psychological experiments from leading social science journals, \citet{cui_can_2024} find that GPT-4 reproduces 76.0\% of main effects, closely aligning with human outcomes in both direction and statistical significance. Similarly, \citet{kirshner_artificial_2024} demonstrate that GPT-4o replicates eight of nine classic findings in operations management, with treatment effects that closely track human behavior in both magnitude and direction. These results suggest that LLMs can reduce the cost and logistical complexity of human-subject experiments while enabling the exploration of counterfactuals and hypothetical conditions at scale \cite{hwang_human_2025}.

While many cases demonstrate that LLMs can approximate human cognitive and behavioral outputs in controlled experiments, it remains unclear whether these models can consistently capture the more nuanced ``noise'' \cite{slifkin_is_1998, de_petrillo_variation_2021, havrilla_understanding_2024} observed in real-world human decision-making. This subtle ``noise'' is essential because it shapes how and when people deviate from normative predictions. This is important information that determines the validity of synthetic-subject replacements in behavioral experiments, and the robustness and fairness of systems that rely on LLM simulations. Without the stochastic fingerprints of actual human decision makers, models risk over-estimating equilibrium convergence or mis-allocating welfare.
For instance, \citet{kitadai_can_2024} show that LLM agents with stronger reasoning abilities tend to produce outcomes closer to theoretical optima than to the actual results observed in human experiments. In another study of altruistic behavior in dictator games, \citet{ma_can_2024} finds that LLMs fail to reproduce the internal deliberation processes underlying human decision-making. These findings suggest that LLM agents may diverge from human subjects’ behavior, underscoring the need for careful evaluation of whether their decision patterns reflect the variability, inconsistency, and adaptive heuristics that characterize human behavior in complex tasks.

\subsection{LLM Behavior Evaluation}

As LLMs are increasingly deployed across domains that require human-level judgment and interaction, evaluating and aligning their behavior has become a central concern in NLP research \cite{wang_aligning_2023, liu_g-eval_2023, yao_instructions_2023}. Traditional evaluation frameworks often emphasize surface-level metrics like factual accuracy, linguistic coherence, or syntactic completeness \cite{yaldiz_not_2025}. Along with the recognition of LLMs' value in human simulations, human-alignment benchmarks such as ``HHH'' (helpful, honest, and harmless) have been introduced to assess normative alignment with intended responses \cite{askell_general_2021}. While these metrics provide important insights into correctness and linguistic quality, they offer limited information about the underlying behavioral processes that govern model decisions.

Recent work has begun to move beyond traditional evaluation metrics by introducing more nuanced frameworks and benchmarks to assess LLM behavior in structured decision contexts. For example, \citet{jia_decision-making_2024} develop a behavioral economics-inspired evaluation framework that quantifies decision patterns like risk preference, probability distortion, and loss aversion. Their results show that LLMs can reproduce some of these behavioral signatures, but their sensitivity to socio-demographic framing often leads to inconsistent patterns. \citet{chen_xplainllm_2024} introduce XplainLLM, a dataset and accompanying explanation framework designed to illuminate LLMs' internal reasoning behavior. In a related line of inquiry, \citet{ross_llm_2024} apply utility theory to analyze economic decision-making by LLMs. They find that while models often produce economically coherent responses in isolated settings, they fail to maintain consistent behavior across varying payoff structures or decision contexts. Furthermore, \citet{macmillan-scott_irrationality_2024} evaluate LLM susceptibility to cognitive biases and reveal that LLMs respond incorrectly in ways that differ from human-like biases. Together, these studies reveal the importance of evaluation methods that go beyond output correctness and instead capture how LLMs simulate plausible human behavior across conditions and frames of reference.

\subsection{Research Gap}

Previous studies show that LLM agents can achieve or even exceed human-level performance targets \citep{chen_put_2024,cai_rtbagent_2025,shah_evidence_2024}.
However, these studies are outcome-oriented, evaluating LLMs mostly on profit or efficiency. They ignore the decision path by which humans reach those outcomes: fluctuating exploration, myopic overreaction, and gradual strategy revision over repeated feedback \citep{slifkin_is_1998,de_petrillo_variation_2021}. 

Our work addresses this gap by proposing a process-oriented evaluation framework with progressive interventions (\emph{Intrinsicality}, \emph{Instruction}, \emph{and Imitation}). We evaluate LLM behaviors on two classic tasks in economics. By comparing LLM behaviors to standard theory and human data, we assess whether LLMs exhibit variability and adaptability in decision-making.

\section{Methodology}

\subsection{Task Description}
\label{sec:task}

We apply our process-oriented framework to two classic decision-making tasks: the \textit{second-price auction} and the \textit{newsvendor problem}. These tasks differ substantially in structure and complexity. The auction task involves strategic reasoning and a discontinuous payoff function, where outcomes depend on the interaction between the reserve price and external bidder valuations. In contrast, the newsvendor task features a smooth, continuous payoff structure and requires threshold-based optimization under cost and demand uncertainty. This contrast enables us to evaluate whether LLMs exhibit consistent behavioral patterns across distinct economic mechanisms.

\textbf{Second-Price Auction}
Our primary experiment is based on the second-price auction mechanism \cite{edelman_internet_2007, cooper_understanding_2008}. In this task, LLM agents act as sellers aiming to maximize total profit over 60 rounds. Following the design of prior human-subject experiments conducted at a U.S. university \cite{davis_auctioneers_2011, davis_replication_2023}, each LLM agent is assigned a unique profile with age, gender, and field of study. Before the auction begins, they receive complete instructions, including the rules of second-price auctions, examples of profit calculation, and illustrations of historical bidder valuation distributions.
For every round \( t \in \{1, \dots, 60\} \), a subject sets a reserve price \( r_t \in \{0, \dots, 100\} \). Each agent is paired with a group of simulated bidder valuations \( \mathbf{b}_t \), sorted in descending order \( b^{(1)}_t \ge b^{(2)}_t \ge \dots \). An item is sold if the highest bid exceeds the reserve price; otherwise, no sale occurs.

In each round, the agent is matched with a group of simulated bidder valuations drawn from one of two known distributions: the \textit{Cube-root distribution}, defined by \( F(v) = \left( \frac{v}{100} \right)^{1/3} \), or the \textit{Cube distribution}, defined by \( F(v) = \left( \frac{v}{100} \right)^3 \). These correspond to left-skewed (\( \mu = 25, \sigma = 28.4 \)) and right-skewed (\( \mu = 75, \sigma = 19.4 \)) settings, respectively, over a common support. The number of bidders per round is randomly chosen from \{1, 4, 7, 10\}. Agents receive feedback on profit after each round and adjust their reserve prices accordingly.

\textbf{Newsvendor Problem}
In this task, LLM agents act as the vendor deciding how many newspapers to order before knowing the actual demand \cite{schweitzer_decision_2000}. In every round \( t \in \{1, \dots, 30\} \), a subject chooses an order quantity \( q \) before observing stochastic demand \( D \sim \mathcal{U}[0, 300] \). The unit cost \( c \) and unit selling price \( p \) vary by round. Agents earn profit according to: \( \text{Profit} = p \cdot \min(q, D) - c \cdot q \). The optimal order quantity \( q^* \) is given by the critical fractile rule: \( q^* = F^{-1}\left( \frac{p - c}{p} \right) \), where \( F \) is the cumulative distribution of demand.

Table~\ref{tab:variable} summarizes the key variables and evaluation metrics. While task-specific variables (e.g., rPrice, order bias) differ, we have common metrics for behavior similarity (Kolmogorov–Smirnov distance) and variability (entropy).

\begin{table}[ht]
\centering
\scriptsize
\renewcommand{\arraystretch}{1.5}
\setlength{\tabcolsep}{7pt}
\begin{tabular}{ll}
\toprule
\textbf{Metric} & \textbf{Definition} \\
\midrule
\textbf{KS distance} &
$D_{\text{KS}} = \sup_x \left| F_{\text{LLM}}(x) - F_{\text{Human}}(x) \right|$ \\
\textbf{Behavioral entropy} &
$H = -\sum_{v} P(a_t = v) \log_2 P(a_t = v)$ \\
\midrule
\multicolumn{2}{l}{\textsc{Auction}}\\ 
\midrule
\textbf{Sale indicator} &
$s_t = \mathbb{1}[\,r_t \le b^{(1)}_t\,]$ \\
\textbf{Profit} &
$p_t = 
\begin{cases}
\max\{r_t,\, b^{(2)}_t\}, & \text{if } s_t = 1 \\
0, & \text{if } s_t = 0
\end{cases}$ \\
\textbf{Sell-through} &
$\text{STR} = \frac{1}{T} \sum_{t=1}^{T} s_t$ \\
\textbf{Premium capture} &
$\text{PCR} = \frac{\sum_{t=1}^{T} \mathbb{1}[\,(s_t = 1) \land (r_t > b^{(2)}_t)\,]}{\sum_{t=1}^{T} s_t}$ \\
\midrule
\multicolumn{2}{l}{\textsc{Newsvendor}}\\ 
\midrule
\textbf{Profit} &
$p_t = \min(q_t, D_t) \cdot p - q_t \cdot c$ \\
\textbf{Order bias} &
$\text{Bias}_t = q_t - q^*_t$ \\
\bottomrule
\end{tabular}
\caption{Key variables and evaluation metrics. For both tasks, $t$ indexes the round, and $T$ is the total number of rounds.  In the auction task, $r_t$ is the reserve price, $b^{(1)}_t$ and $b^{(2)}_t$ are the highest and second-highest bidder valuations in round $t$. In the newsvendor task, $q_t$ is the quantity ordered, $D_t$ is realized demand, $p$ is the unit selling price, $c$ is the unit cost, and $q^*_t$ is the theoretical optimal quantity. The action $a_t$ refers to the subject's decision at round $t$ (e.g., $r_t$ or $q_t$). KS distance measures distributional similarity between LLM and human decisions, and entropy quantifies behavioral variability.}

\label{tab:variable}
\end{table}

Empirical studies show that humans systematically deviate from optimal strategies in both settings. In auctions, they tend to increase reserve prices as the number of bidders increases, a pattern linked to bounded rationality, overconfidence, or heuristic beliefs about competition. In the newsvendor task, over-ordering in low-margin conditions and demand-chasing behavior are commonly observed. By comparing LLM behavior to both theoretical optima and human benchmarks across these two tasks, we assess whether LLMs conform to normative expectations and whether they present key patterns in human decision-making.

\subsection{Setup}
\label{sec:model_setup}

We instantiate LLM agents using state-of-the-art models: GPT-4o \cite{openai_gpt-4o_2024}, Claude 3.5 Sonnet \cite{anthropic_introducing_2024}, and Claude 3.7 Sonnet \cite{anthropic_claude_2025}. For each model, we simulate 40 agents, each assigned a unique profile constructed from real demographic data (gender, race, age, and academic background). Before each task, agents receive complete instructions, including the rules of the auction or newsvendor problem, along with illustrative examples of profit calculation. Full instruction texts are provided in Appendix~\ref{appendix:exp_text}.

To systematically evaluate LLM behavior, we propose the evaluation framework with progressive interventions: (1) \textbf{Intrinsicality}, where agents complete the task identically to human subjects, without any intervention; (2) \textbf{Instruction}, where agents receive additional framing about risk preferences (e.g., risk-seeking or risk-averse); and (3) \textbf{Imitation}, where human behavioral traces are provided as in-context examples.
In the imitation condition, we examine how LLMs respond when exposed to human noise and behavioral variance through partial demonstrations. Agents are given the historical records of real human data, including reserve prices, profits, and bidder valuations (auction), or order quantities, profits, and demands (newsvendor). Then, they complete the remaining rounds under three conditions: (1) \textbf{Direct Imitation}: Replicate the humans' behavior patterns as closely as possible; (2) \textbf{Context-Aware Imitation}: Use the human history as a reference while independently selecting actions that aim to maximize profit; (3) \textbf{Theory-Guided Imitation}: Given an explanation of the normative solution, use human history as context while optimizing for profit.

Task parameters such as bidder valuations and demand distributions are held constant across agents with matching demographic profiles, ensuring consistency with the conditions used in the original human-subject studies. \emph{Our prompt design for interventions is provided in Appendix~\ref{appendix:intervention}.}

\subsection{Experimental Controls and Robustness}
\label{sec:robustness}

To ensure the reliability and reproducibility of our findings in the main and supplement tasks, we conducted each experimental configuration three times per LLM and intervention condition. We assessed consistency by computing pairwise Pearson correlations of key behavioral sequences (e.g., reserve prices in the auction, order quantities in the newsvendor task) and profit outcomes, observing correlations exceeding 0.99 across runs. Consequently, we report results from the first iteration throughout the paper.
We performed robustness checks across different temperature settings (\texttt{0.0}, \texttt{0.3}, \texttt{0.7}, \texttt{1.0}), finding that both reserve price trajectories and profit distributions remained consistent. Thus, we adopt a temperature of 1.0 as the default setting for all reported experiments.
For comparisons such as total profit or KS distance, we conducted t-tests and reported corresponding $p$-values. These tests confirm that our observed differences are statistically significant and not driven by random variation across agent populations or experimental replications.

\section{Main Experiment (Auction)}
\subsection{Intrinsicality}
\label{sec:intrinsicality}

In the intrinsicality experiment, we follow an identical experimental process for human subjects and LLM agents, aiming to show LLM agents' default behavior patterns.

\textbf{Behavioral Consistency}
Table~\ref{tab:main_summary_stats} presents summary statistics for reserve prices and profit across all subjects and rounds under the intrinsicality condition. We observe that LLM agents achieve similar profits. Meanwhile, their reserve price distributions vary substantially. Human subjects set the highest reserve prices on average ($\mu$=27.3, $\sigma$=23.2), reflecting greater variability than LLM agents.
\begin{table}[htbp]
\small
\begin{tabular}{p{1.4cm}lccccc}
\toprule
\textbf{Model} & \textbf{Mean} & \textbf{SD} & \textbf{Min} & \textbf{Max} & \textbf{S} \\
\midrule
\multicolumn{6}{l}{\textit{Profit per round (across 40 subjects × 60 rounds)}} \\
Claude 3.5  & 34.54 & 28.35 & 0.0 & 97.0 & 4.79 \\
Claude 3.7  & 34.13 & 28.38 & 0.0 & 97.0 & 5.00 \\
GPT-4o      & 33.00 & 29.45 & 0.0 & 97.0 & 4.19 \\
Human       & 32.83 & 30.37 & 0.0 & 99.0 & 5.06 \\
\midrule
\multicolumn{6}{l}{\textit{rPrice per round (across 40 subjects × 60 rounds)}} \\
Claude 3.5  & 23.37 & 8.44  & 1.0 & 75.0 & 2.92 \\
Claude 3.7  & 18.22 & 7.60  & 0.0 & 60.0 & 2.96 \\
GPT-4o      & 16.12 & 11.68 & 0.0 & 80.0 & 1.43 \\
Human       & 27.31 & 23.25 & 0.0 & 100.0 & 5.08 \\
\bottomrule
\end{tabular}
\caption{Summary statistics of profit and reserve prices under no intervention (Intrinsicality). Each source contributes 2,400 data points (40 subjects \( \times \) 60 rounds), with metrics computed at the round level. \emph{S} represents entropy.}
\label{tab:main_summary_stats}
\end{table}

The inconsistency between reserve price and profit is expected given the mechanics of second-price auctions: Final profit is primarily influenced by the bidder valuation distribution rather than the reserve price, as long as the reserve price is not set prohibitively high. In this context, LLM agents, which tend to set lower reserve prices than humans, still achieve comparable profits because their items are more likely to sell, and competitive bidding drives up the final transaction price. Notably, we observe a ``threshold phenomenon'' in human behavior, where some subjects push reserve prices to the extremes of the allowable range. In contrast, LLM agents exhibit a more constrained price range, not exceeding 80.

Overall, LLM agents outperform human subjects in maximizing total profit ($p<0.05$). The lower intra-agent and inter-agent variance in LLM decisions reflects a fundamental behavioral divergence from humans, suggesting that current models optimize effectively but lack the variations that characterize human strategic reasoning.

\textbf{Strategic Preferences}

\begin{figure}[htbp]
\centering
\includegraphics[width=0.45\textwidth]{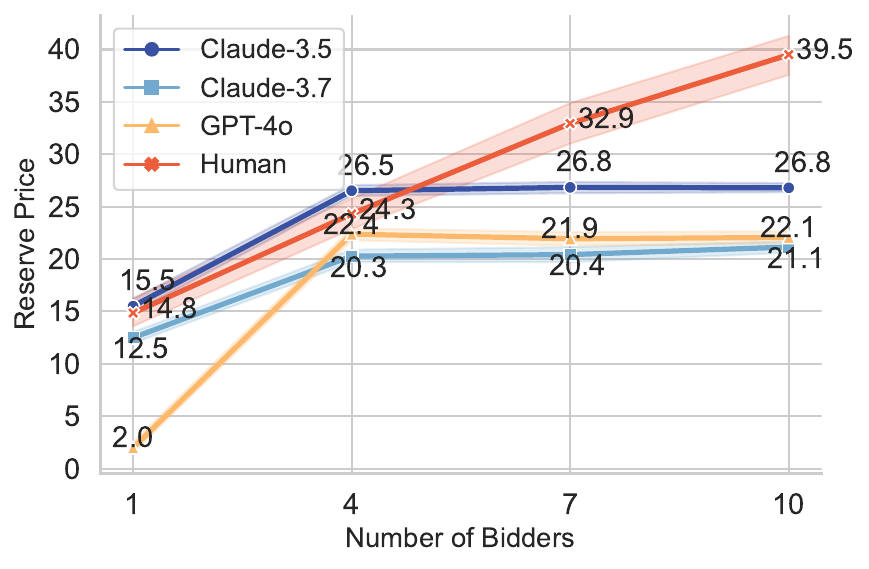}
\caption{
Reserve price variation with the number of bidders, aggregated across 40 agents per bidder-number group. Each line represents the mean reserve price, with shaded areas indicating 95\% confidence intervals.}
\label{fig:rprice_line}
\end{figure}

 To investigate human subjects and LLM agents' pricing strategies, we group 2,400 subject-level reserve price observations by the number of bidders. We average the reserve price to capture the overall strategic response to varying levels of market competition. The results are presented in Figure~\ref{fig:rprice_line}.

As the number of bidders increases from one to four, all LLM agents raise their reserve prices. Claude-based agents align more closely with human behavior, starting near 15 and rising to about 25. GPT-4o is more conservative, setting 2 on average with one bidder. Beyond four bidders, LLM strategies converge and stabilize between 20 and 26, while humans continue increasing their prices, reaching nearly 40 with ten bidders.

\begin{figure}[htbp]
\centering
\includegraphics[width=0.45\textwidth]{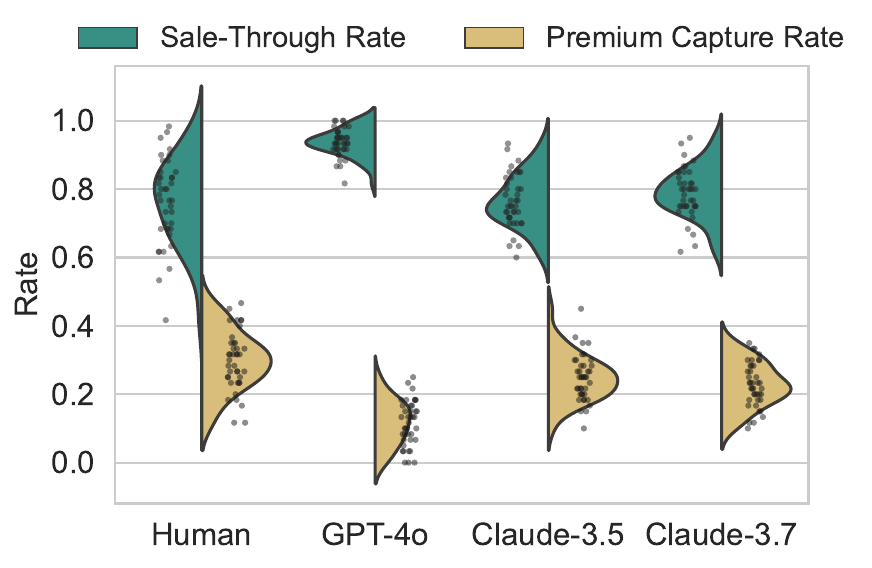}
\caption{
Split violin and dot plot comparing STR and PCR. The half-violins indicate variability within each metric, and the dots show 40-subject values.}
\label{fig:sale_premium_rate}
\end{figure}

To learn more about selling tactics, we examine the sell-through rate (STR) and the premium capture rate (PCR). For each source, we collect 40 subject-level observations and visualize their distributions in Figure~\ref{fig:sale_premium_rate}, highlighting distinct behavioral patterns across human and LLM agents.

All LLM agents have high STR with a median around 0.7, indicating a general preference for securing the sales. Meanwhile, they exhibit lower and less variable premium capture rates compared to human subjects, suggesting a tendency to prioritize transaction completion over profit maximization. Notably, GPT-4o stands out with an unusually high and tightly clustered sell-through rate near 0.9, alongside a significantly lower PCR concentrated around 0.1. This pattern indicates a particularly conservative pricing strategy.

Overall, in the absence of intervention, LLM agents diverge from the adaptive and heterogeneous nature of human decision-making. In the following sections, we introduce risk-framing instruction and imitation strategies. By comparing to the intrinsic results, we assess how these interventions can impact LLM behavior.

\subsection{Instruction}
\label{sec:risk}

Risk preference is a critical factor influencing human behavior in auction settings \cite{myerson_optimal_1981, cooper_understanding_2008}. To examine whether LLM agents exhibit similar sensitivity, we introduce risk-framed instructions, risk-averse and risk-seeking, into the original experiment and evaluate how they adjust their pricing strategies. We introduce the Intrinsicality results from Section~\ref{sec:intrinsicality} for reference.

\begin{figure}[htbp]
\centering
\includegraphics[width=0.45\textwidth]{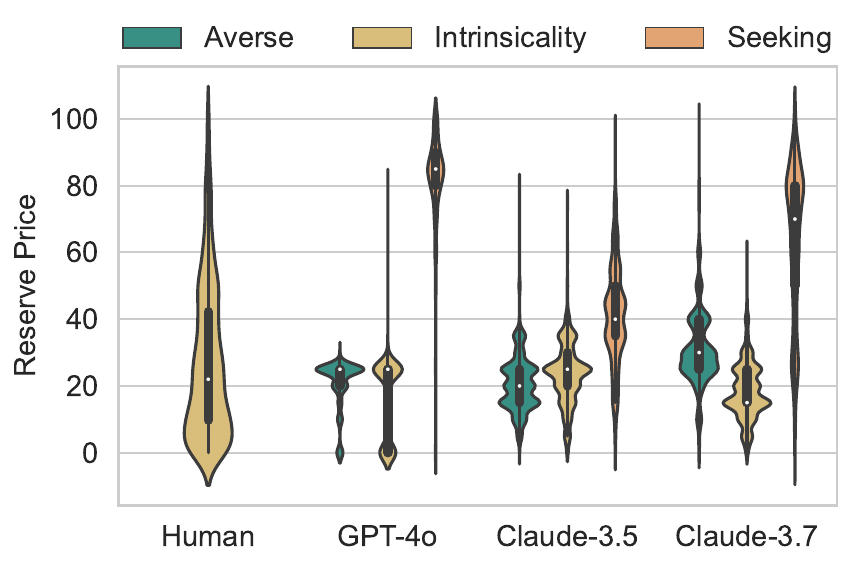}
\caption{
Reserve price distributions across risk-framed instruction conditions.}
\label{fig:risk_violin}
\end{figure}

Figure~\ref{fig:risk_violin} presents the reserve price distributions across risk-framing instructions. Risk-seeking instructions elicit higher reserve prices from LLM agents. Interestingly, reserve prices under risk-averse instructions are closely aligned with those without any intervention (Intrinsicality). This suggests that, by default, LLM agents tend to adopt a conservative strategy in the absence of explicit instruction. Overall, LLM agents produce tightly clustered and coherent distributions within each condition, indicating strong internal alignment to instruction. While LLMs flexibly adapt to risk framing, they still exhibit less behavioral diversity than human subjects.

\subsection{Imitation}
\label{sec:imitation}

We further examine how LLM agents respond when real human noise is injected via in-context learning. By providing the first 30 rounds of human data (including reserve prices, profits, and bidder valuations), we task LLM agents with completing the remaining 30 rounds under conditions: Direct imitation (Direct), Context-aware imitation (Context), and Theory-guided imitation (Theory).
For reference, we include results from Intrinsicality from Section~\ref{sec:intrinsicality}, reflecting LLM agents' intrinsic behaviors.

\begin{figure*}[htbp]
  \centering
  \begin{subfigure}[t]{0.32\textwidth}
    \includegraphics[width=\linewidth]{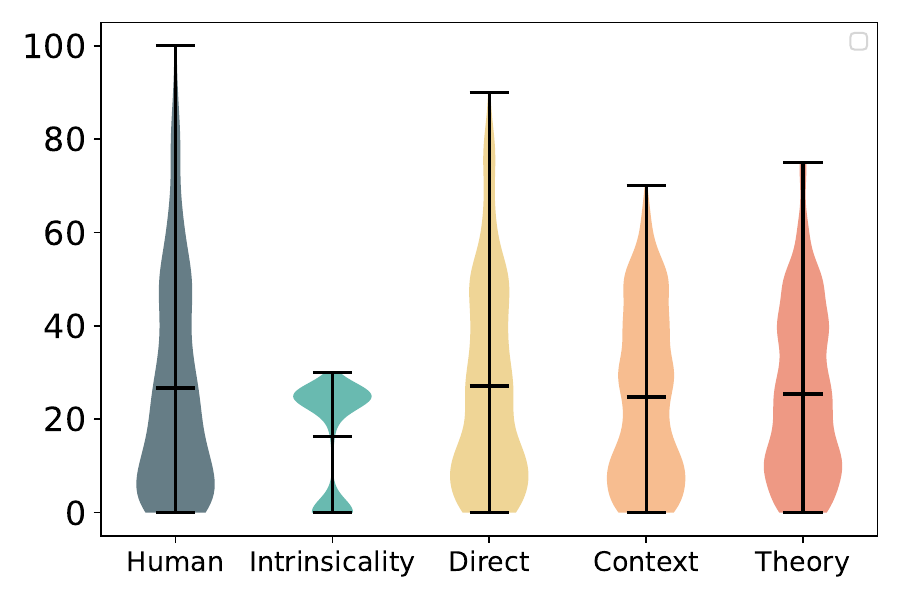}
    \caption{Reserve price distribution}
    \label{fig:q3_reserve_distribution}
  \end{subfigure}
  \hfill
  \begin{subfigure}[t]{0.32\textwidth}
    \includegraphics[width=\linewidth]{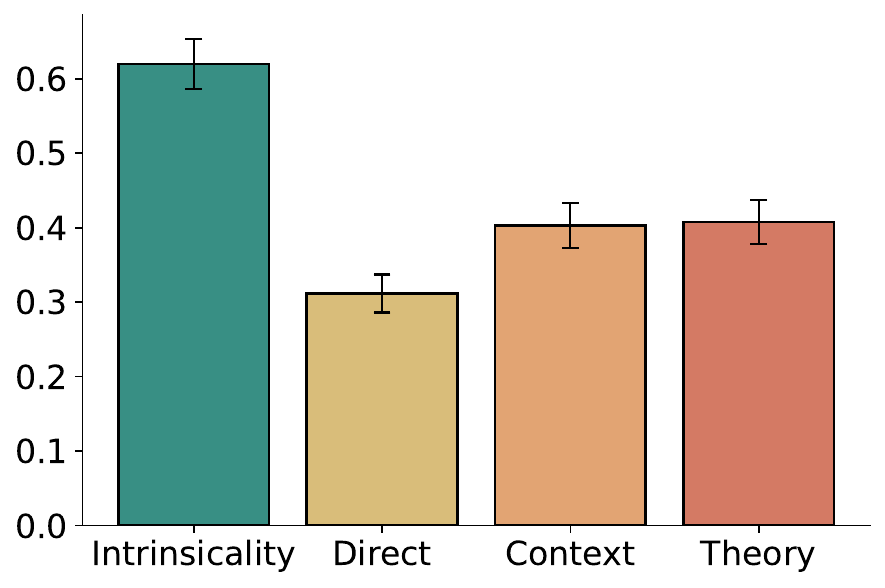}
    \caption{KS distance to human}
    \label{fig:q3_ks}
  \end{subfigure}
  \hfill
  \begin{subfigure}[t]{0.32\textwidth}
    \includegraphics[width=\linewidth]{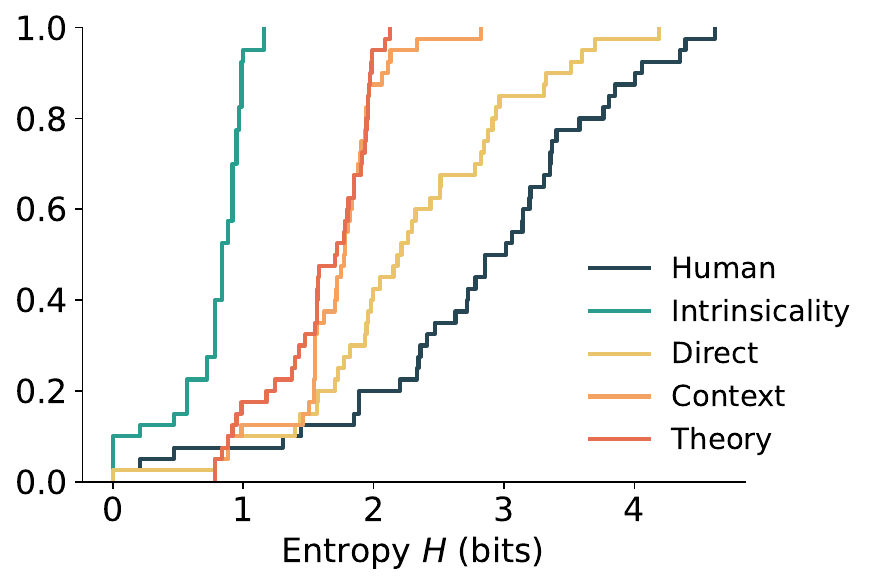}
    \caption{Behavioral entropy}
    \label{fig:q3_entropy}
  \end{subfigure}
  \caption{Comparison across imitation modes based on outcome, behavioral fidelity, and entropy.}
  \label{fig:q3}
\end{figure*}

Figure~\ref{fig:q3_reserve_distribution} compares the distribution of reserve prices under imitation conditions to human and intrinsic results. Without intervention (intrinsicality), LLM agents exhibit narrow and concentrated reserve price ranges, reflecting deterministic optimization. Direct imitation recovers much of the dispersion present in human decisions, while context-aware and theory-guided imitations also exhibit intermediate variability.

Figure~\ref{fig:q3_ks} presents the average Kolmogorov-Smirnov distance between 40 LLM agents' reserve price trace and their paired human subjects over rounds 31 to 60. Direct imitation most closely resembles human behavior (\(\overline{D_{\text{KS}}} \approx 0.31\)). In contrast, context-aware (\(\overline{D_{\text{KS}}} \approx 0.40\)) and theory-guided imitations (\(\overline{D_{\text{KS}}} \approx 0.40\)) show slightly greater divergence. Intrinsically, LLMs deviate most ($p<0.05$), with a nearly doubled distance (\(\overline{D_{\text{KS}}} \approx 0.62\)).  Despite being encouraged for independent reasoning, even context-aware and theory-guided imitation tend to lean toward direct imitation ($p>0.05$). In other words, human data can override the LLM agents' default behavior or the guidance of optimal theory.

Figure~\ref{fig:q3_entropy} shows the ECDF of reserve price entropy, capturing the diversity of price-setting strategies. Human subjects are most stochastic: only 25\% fall below \(H = 2\,\mathrm{bits}\). Under Intrinsicality, LLM agents rarely exceed \(1.2\,\mathrm{bits}\). Direct imitation roughly doubles this diversity, and both context-aware and theory-guided imitations also increase entropy, though none approach the upper bounds of human variability (\(H > 4\,\mathrm{bits}\)).

Overall, LLM agents can be effectively influenced by the different imitations towards human subjects. However, even the best direct imitation exhibits less noise and variability than humans, suggesting that achieving full behavioral realism may require additional interventions.

\section{Supplement Experiment (Newsvendor)}
\label{appendix:newsvendor_results}

We apply the same evaluation framework to the newsvendor task to assess its generalizability. Across all interventions, LLM behavior again departs from human patterns, most notably in its consistent ordering strategies.

Under \textbf{intrinsicality}, LLMs produce highly concentrated order quantities near the theoretical optimum \( q^* \), contrasting with the wide and variable decisions of human subjects. With the \textbf{instruction}, LLMs adjust as expected, and LLMs are sensitive to explicit risk framing, though their decisions remain more stable than human subjects. For \textbf{imitation}, LLMs present the most human-like behavior with direct imitation. Context-aware and theory-guided imitations achieve better profit alignment but diverge more from humans.

These results are similar to the auction task: LLMs behave deterministically by default but can be impacted toward more human-like behavior with appropriate interventions. Full quantitative results are provided in Appendix~\ref{appendix:newsvendor_results}.

\section{Discussion}
\label{sec:discussion}

Our findings reveal that LLM agents inherently converge toward rational, profit-maximizing behaviors that diverge from the noisy and variable patterns typical of human decision-making. One likely explanation is that LLMs are trained to minimize predictive loss over large-scale corpora, which promotes high-probability, low-variance outputs \cite{lee_local_2023}. Additionally, greedy decoding or low-temperature sampling in tasks \cite{tang_top-nsigma_2024} further suppresses behavioral variability, reinforcing deterministic responses.

Our evaluation shows that textual framing and in-context demonstrations can shift LLM behavior toward more human-like patterns. However, none of these interventions fully reproduces the heterogeneity intrinsic to human decision-making. Moreover, injecting human data introduces the risk of importing biases, potentially undermining the rational consistency often regarded as a strength of LLMs \cite{havrilla_glore_2024, wang_rethinking_2024}.

A key implication of our study is that employing LLMs as proxies for human subjects should report behavioral audits alongside experimental results. Where gaps in variability persist, they must be acknowledged and contextualized to assess the credibility of LLMs as substitutes for human decision-makers. This transparency is essential for evaluating the appropriateness of LLMs in behavioral research and for informing future model development that balances optimization with the stochasticity that gives behavioral data its explanatory value, especially in synthetic social-science experiments where variability is an important signal.

\section{Conclusion}
\label{sec:conclusion}
This study presents a process-oriented framework to evaluate the behavioral fidelity of LLM agents in dynamic decision-making tasks. Across two economic experiments (second-price auction and the newsvendor problem), we find that LLMs consistently adopt low-variance rational strategies that diverge from human behavior in both variability and adaptability. While risk framing and imitation can partially nudge models toward human-like behavior, these interventions fall short of reproducing the stochastic and context-sensitive decision patterns observed in human subjects. Our results emphasize the necessity for more process-aware evaluation in behavioral applications of LLMs and offer a promising method for auditing their suitability as synthetic human proxies in social science research.

\section*{Limitations and Future Work}
\label{sec:limitations}

We acknowledge several limitations in this study.

First, although we evaluate LLM behavior across two classic decision-making settings, both are single-agent settings with the profit-optimization target. Extending the framework to more interactive, multi-agent environments, such as bargaining, coordination, or deception games, would offer a richer understanding. 

Second, our interventions are limited to static, text-based inputs. These approaches may not fully capture or induce the complexity of human-like variability. Future work could explore more dynamic agent conditioning methods, including multi-turn interactions, memory-based adaptation, or reinforcement fine-tuning. 

Finally, while we benchmark against high-quality empirical human data, the underlying human experiments are based on a specific participant pool. Replicating these comparisons across more diverse populations would strengthen claims about generalizability.

\section*{Ethical Considerations}
This study incorporates human data in behavioral economics obtained from prior work conducted under institutional review board (IRB) approval. No new human data were collected. We faithfully reproduce participant profiles (e.g., age, gender, academic background) from the original studies to preserve the integrity of agent-level comparisons. 
We acknowledge the risks of over-interpreting LLM outputs as reflective of human cognition. While our work explores whether LLMs replicate certain behavioral patterns, we avoid reinforcing stereotypes or drawing normative conclusions from demographic attributes. We release no personally identifiable data and adhere to ethical standards regarding simulation fidelity, model transparency, and responsible reporting of results.

\bibliography{references}

\appendix

\section{Supplementary Results (Newsvendor)}
\label{sec:newsvendor_results}

To assess the generalizability of our framework beyond the main experiment, we apply the same evaluation procedure to the newsvendor problem, a single-agent inventory task under demand uncertainty. As in the auction setting, we examine LLM behavior under intrinsicality, instruction, and imitation conditions.

\begin{figure}[htbp]
\centering
\includegraphics[width=0.45\textwidth]{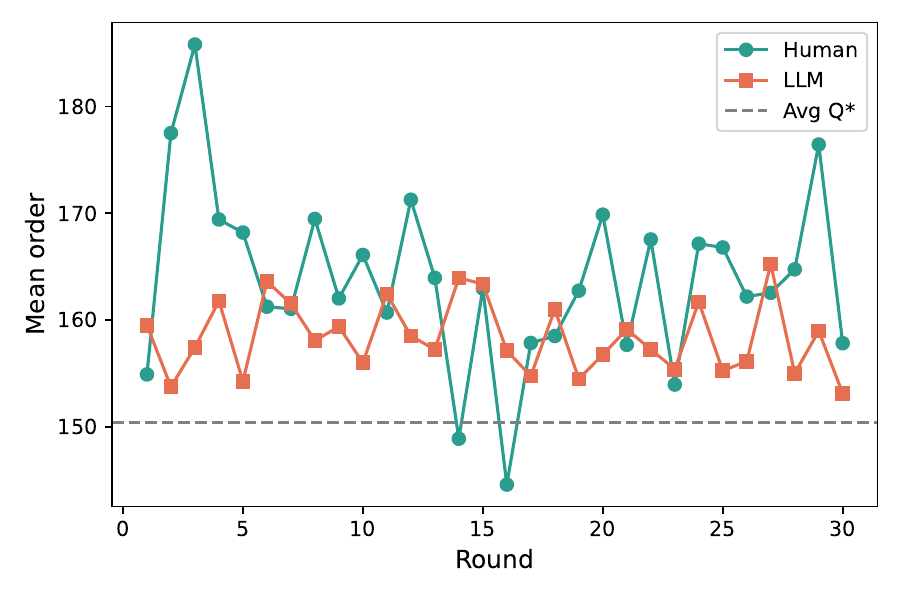}
\caption{Order quantities over rounds: Human vs. intrinsic LLM agents.}
\label{fig:newsvendor_human_vs_intrinsic}
\end{figure}

\paragraph{Intrinsicality}  
Figure~\ref{fig:newsvendor_human_vs_intrinsic} shows the average order quantities over time. Human subjects display substantial round-to-round variability and frequently deviate from the optimal order quantity \( q^* \). In contrast, intrinsic LLM agents consistently produce tightly clustered orders near \( q^* \), indicating a low-variance, profit-oriented strategy. While effective in terms of aggregate performance, this consistency lacks the adaptive fluctuations commonly observed in human behavior.

\begin{figure}[htbp]
\centering
\includegraphics[width=0.45\textwidth]{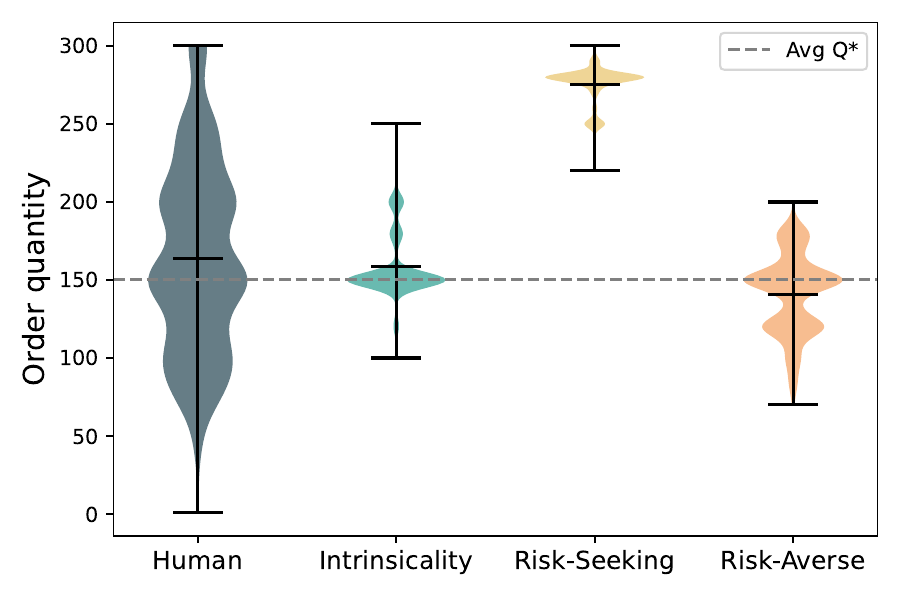}
\caption{Order quantity distributions across risk-framed instruction conditions.}
\label{fig:newsvendor_instruction_order_distribution}
\end{figure}

\paragraph{Instruction}  
Under risk-framed prompts (Figure~\ref{fig:newsvendor_instruction_order_distribution}), LLM agents respond directionally as expected: risk-seeking instructions lead to higher order quantities, while risk-averse instructions induce lower orders. These shifts are aligned with human interpretations of risk and indicate that LLMs are responsive to abstract risk framing. Nevertheless, the resulting distributions remain narrower than those of human subjects, further reinforcing the tendency toward reduced behavioral variability.

\begin{figure*}[htbp]
  \centering
  \begin{subfigure}[t]{0.32\textwidth}
    \includegraphics[width=\linewidth]{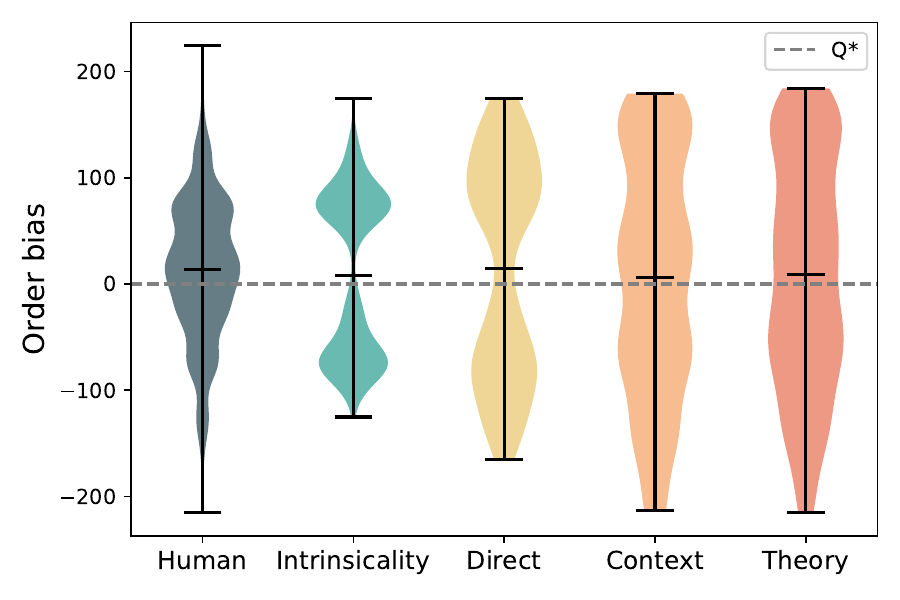}
    \caption{Order bias distribution}
    \label{fig:newsvendor_order_bias_distribution}
  \end{subfigure}
  \hfill
  \begin{subfigure}[t]{0.32\textwidth}
    \includegraphics[width=\linewidth]{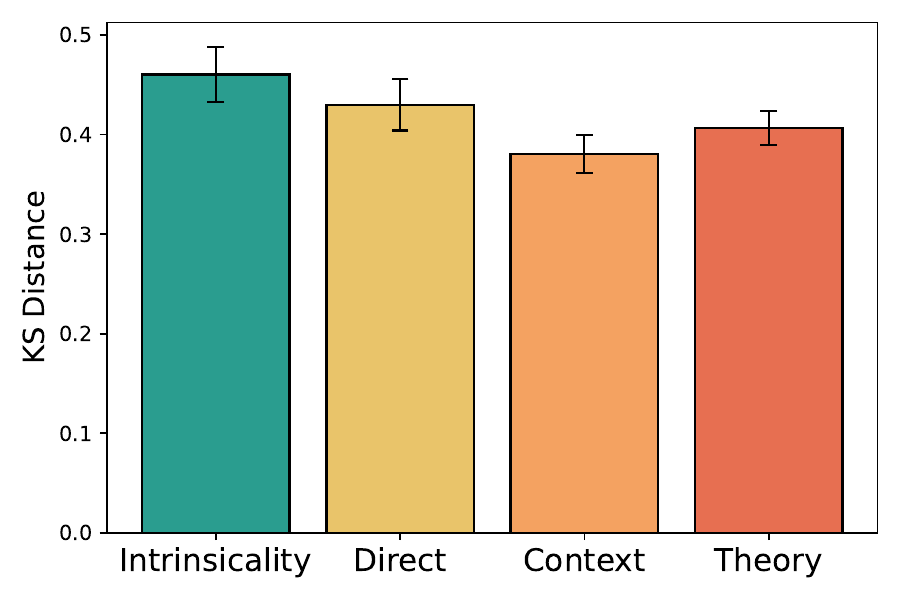}
    \caption{KS distance to human}
    \label{fig:newsvendor_ks_distance}
  \end{subfigure}
  \hfill
  \begin{subfigure}[t]{0.32\textwidth}
    \includegraphics[width=\linewidth]{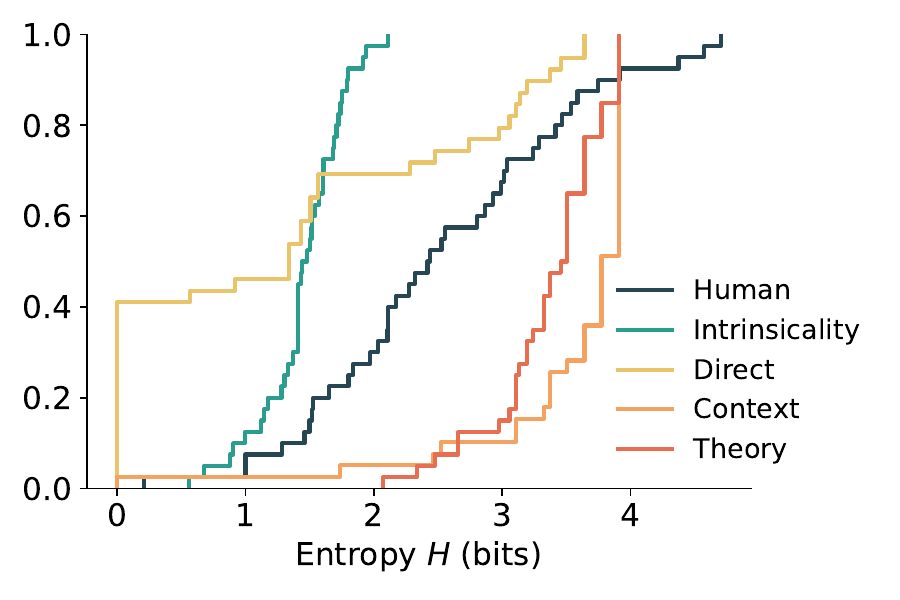}
    \caption{Behavioral entropy}
    \label{fig:newsvendor_entropy_ecdf}
  \end{subfigure}
  \caption{Comparison across imitation modes based on outcome alignment, distributional similarity, and entropy.}
  \label{fig:newsvender_imitation}
\end{figure*}

\paragraph{Imitation}  
We further evaluate three imitation prompting conditions. As shown in Figure~\ref{fig:newsvendor_order_bias_distribution}, direct imitation yields the most human-like order dispersion, partially recovering the variability observed in human subjects. In contrast, LLMs in context-aware and theory-guided conditions converge on more stable and performance-driven strategies with reduced bias. This trade-off is evident in both KS distance (Figure~\ref{fig:newsvendor_ks_distance}) and behavioral entropy (Figure~\ref{fig:newsvendor_entropy_ecdf}), where direct imitation achieves the closest match to human behavior, while intrinsic and theory-guided responses remain more deterministic.

Across all conditions, LLM agents in the newsvendor task exhibit behavioral trends similar to those observed in the auction experiment: they default to consistent, low-variance strategies but can be guided toward more human-like patterns through framing or demonstration. Even so, imitation does not fully recover the variability inherent to human decision-making. These results reinforce the main findings of our study and further support the validity of our evaluation framework. They also highlight the importance of careful behavioral auditing when considering LLMs as substitutes for human subjects in social science research.

\section{Second-Price Auction Instruction}
\label{appendix:exp_text}

\subsection{Overview}
You are a seller of a fictitious product. To make money, you must sell that product in an auction. In each auction, you will be selling your product to a different number of computerized buyers. To sell the product, you enter the minimum value that you are willing to sell this product for; this amount is your \textbf{reserve price}.

Before entering your reserve price, you will know the number of computerized buyers participating in your auction for each round. Each of these buyers will have a maximum willingness to pay for your product. The maximum willingness to pay is an integer between 0 and 100.

In each auction, bids ascend until all buyers reach their maximum willingness to pay. Some buyers will be forced to stop bidding as their limits are reached. The auction ends when one buyer remains. The amount where each buyer stops bidding in the auction is called the \textbf{Drop-Out Price}.

The buyer with the highest Drop-Out Price wins so long as that amount is equal to or above your reserve price. If all of the Drop-Out Prices are below your reserve price, your product will not be sold, and you will earn zero in that round.

\subsection*{Profit Calculation}
\begin{itemize}
    \item Your Profit = Second Highest Drop-Out Price, if the second highest Drop-Out Price is equal to or above your reserve price.
    \item Your Profit = reserve price, if the highest Drop-Out Price is above the reserve price, but the second highest Drop-Out Price is below the reserve price.
    \item Your Profit = 0, if all the Drop-Out Prices are below the reserve price.
\end{itemize}

Each auction round, you will have one unit of the product to sell, regardless of whether or not you sold it in the previous round. You will participate in 60 consecutive auctions.

For each round, the number of buyers is randomly drawn as 1, 4, 7, or 10, each number being equally likely. The maximum amount a buyer is willing to pay is a whole number from 0 to 100.

\subsection{Examples}
\begin{itemize}
    \item \textbf{Example 1:} If you enter a reserve price of 60, and the second highest Drop-Out Price is 75, you earn a profit of 75 in this round.
    \item \textbf{Example 2:} If you enter a reserve price of 40, and the second highest Drop-Out Price is 30, but the highest Drop-Out Price is 60, you earn a profit of 40 in this round.
    \item \textbf{Example 3:} If you enter a reserve price of 60, and the highest willingness to pay is 55, then none of the buyers have a Drop-Out Price equal to or higher than 60. No bids will be entered, and you earn a profit of 0 in this round.
\end{itemize}

\subsection{Mechanics of Entering a Reserve Price}
Once you output a reserve price, the auction will proceed.

\subsection{Information Displayed After Each Round}
After each round, you will be shown the results. We will show what the Drop-Out Prices were for each buyer or whether the buyer was unable to bid because their maximum willingness to pay was below the reserve price.

You will also see the following information:
\begin{itemize}
    \item The Period
    \item The Reserve price
    \item The Number of Buyers
    \item The Winning Bid
    \item Your Profit for the round
\end{itemize}

You will also see this information for all previous rounds.

\subsection{Sample Drop-Out Prices Table (100 Rounds)}
\begin{center}
\small
\begin{tabular}{cccccccccc}
\toprule
4 & 32 & 0 & 16 & 67 & 47 & 0 & 12 & 0 & 0 \\
1 & 5 & 39 & 0 & 21 & 81 & 5 & 0 & 7 & 1 \\
2 & 1 & 1 & 7 & 3 & 4 & 1 & 84 & 8 & 58 \\
2 & 0 & 62 & 2 & 1 & 0 & 68 & 20 & 92 & 8 \\
5 & 0 & 4 & 0 & 16 & 43 & 1 & 50 & 2 & 0 \\
43 & 7 & 39 & 7 & 6 & 0 & 23 & 25 & 14 & 12 \\
71 & 0 & 17 & 44 & 15 & 15 & 3 & 0 & 84 & 54 \\
1 & 91 & 11 & 60 & 1 & 36 & 91 & 30 & 3 & 0 \\
1 & 1 & 0 & 15 & 0 & 57 & 3 & 9 & 93 & 5 \\
10 & 2 & 18 & 71 & 0 & 12 & 79 & 64 & 19 & 10 \\
\bottomrule
\end{tabular}
\end{center}

\subsection{Drop-Out Price Sample Distribution}
\textbf{Chart Title:} Drop-Out-Price Sample Distribution

\textbf{X-axis:} Drop-Out-Price (10 price ranges)

\textbf{Y-axis:} Frequency (percentage of rounds)

\begin{itemize}
    \item The 0--10 range is the tallest bar, representing approximately 45\% of the total rounds.
    \item The 11--20 range is about 15\%.
    \item The ranges 21--30, 31--40, and 41--50 each account for between 5\% and 10\%.
    \item The 51--60 and higher ranges are each below 5\%.
    \item The 91--100 range is the shortest bar, representing the lowest frequency.
\end{itemize}

\textbf{Below the graph:}
\begin{itemize}
    \item Average: 25
    \item Min: 0
    \item Max: 100
\end{itemize}

\section{Interventions}
\label{appendix:intervention}

This appendix documents the full prompting setup used across both experimental tasks: the second-price auction and the newsvendor problem. We organize all prompt templates under the three levels of intervention defined in our framework: \textbf{Intrinsicality}, \textbf{Instruction}, and \textbf{Imitation}. Each block corresponds to a system or user message provided to the LLM agent, using placeholders (e.g., \texttt{\{round\}}, \texttt{\{price\}}) that were dynamically populated at runtime.

\subsection{Second-Price Auction}

\subsubsection{Intrinsicality}
\begin{lstlisting}[caption={Auction: Intrinsicality (System Prompt)}]
You are an undergraduate student at the University of Michigan.
You are {age}, {gender}, {race}, and studying {program}.

You are about to participate in an experiment in the economics of decision making.

Here are the experiment instructions:
{experiment_instructions}

IMPORTANT:
- Try to maximize your total profit over 60 rounds.
- You can only respond with an integer between 0 and 100 representing the reservation price.
- Do not provide any explanation or additional text in your response.
\end{lstlisting}

\begin{lstlisting}[caption={Auction: Intrinsicality (User Prompt)}]
Here is your last round result:
{last_round_info}

Here is the history of all previous rounds (Drop-Out Prices provides the bidding values from all bidders, and 'None' represents a bid lower than your reserve price):
{history}

Now it's round {current_round}.
Number of Bidders in this round: {current_num_bidders}

What reserve price do you set for this round?
\end{lstlisting}

\subsubsection{Instruction}
\begin{lstlisting}[caption={Auction: Instruction (System Prompt Modifier)}]
You are a risk-seeking decision maker, prioritizing higher-risk reservation prices for the potential of higher profit.
(OR)
You are a risk-averse decision maker, prioritizing lower-risk reservation prices to ensure positive profits.
\end{lstlisting}

\subsubsection{Imitation}
\begin{lstlisting}[caption={Auction: Imitation (System Prompt)}]
You are participating in a 60-round second-price auction experiment.

## Experiment Instructions:
{experiment_instructions}

## Task:
{task}

## Output Format:
Provide your responses in the following format, without any additional text or explanations:

round 31: [reserve_price]
...
round 60: [reserve_price]
\end{lstlisting}

\begin{lstlisting}[caption={Auction: Imitation (User Prompt)}]
## Participant's Auction Results (Rounds 1-15):
{first_auction_result_texts}

## Bidder Information for Prediction (Rounds 31-60):
{last_bid_information_texts}
\end{lstlisting}

\begin{lstlisting}[caption={Auction: Imitation Task Variants}]
Direct Imitation:
1. Review the participant's bidding results from the first 15 rounds.
2. For rounds 30 to 60, replicate the participant's strategy as closely as possible.

Context-Aware Optimization:
1. Review the participant's bidding results from the first 30 rounds as the reference.
2. For rounds 31 to 60, choose reserve prices that you believe will maximize total profit.
3. While the participant's results may inform your strategy, you are free to optimize based on your own reasoning.

Theory-Guided Prediction:
1. Standard auction theory indicates that there is an optimal reserve price that maximizes total profit, regardless of the number of bidders.
2. Review the participant's behavior in the first 30 rounds.
3. For rounds 31 to 60, choose reserve prices that you believe will maximize total profit, using both theory and participant history.
\end{lstlisting}

\subsection{Newsvendor}

\subsubsection{Intrinsicality}
\begin{lstlisting}[caption={Newsvendor: Intrinsicality (System Prompt)}]
You are participating in an inventory management simulation.
In each round, you will decide how many units of a product to order before the selling season begins.
The demand for the product is uncertain but follows a known distribution.
Your objective is to maximize your profit over the course of the simulation.
Your output should be an integer between 0 and 300.
\end{lstlisting}

\begin{lstlisting}[caption={Newsvendor: Intrinsicality (User Prompt)}]
Round {round}
- Selling Price per Unit: {price} USD
- Cost per Unit: {cost} USD
Please enter the number of units you would like to order for this round.
\end{lstlisting}

\subsubsection{Instruction}
\begin{lstlisting}[caption={Newsvendor: Instruction (System Prompt Modifier)}]
Risk-Seeking:
As a risk-seeking manager, you are willing to take chances.
You prefer to over-order in hopes of capturing high sales, even if it means risking unsold inventory.

Risk-Averse:
As a risk-averse manager, you are cautious.
You prefer to under-order to avoid the risk of unsold inventory, even if it means missing some potential sales.
\end{lstlisting}

\subsubsection{Imitation}
\begin{lstlisting}[caption={Newsvendor: Imitation (System Prompt)}]
You are an autonomous agent in a 30-round inventory management experiment.

## Instructions:
In each round, you decide how many units to order before demand is realized.
Demand ranges from 1 to 300 units. Your objective is to maximize profit.
Price and cost vary by round.

## Task:
{task}

## Output Format:
Please respond using this format, one per line:

round 16: [order]
...
round 30: [order]
\end{lstlisting}

\begin{lstlisting}[caption={Newsvendor: Imitation (User Prompt)}]
## Participant History (Rounds 1-15):
{context_text}

## Demand & Pricing Info (Rounds 16-30):
{future_demand_text}
\end{lstlisting}

\begin{lstlisting}[caption={Newsvendor: Imitation Task Variants}]
Direct Imitation:
1. Review the participant's inventory ordering decisions from the first 15 rounds.
2. For rounds 16 to 30, continue their strategy by predicting order quantities that match their decisions.

Context-Aware Optimization:
1. Review the participant's order history from the first 15 rounds as context.
2. For rounds 16 to 30, choose order quantities that you believe will maximize total profit.
3. You may consider the participant's behavior, but are free to reason independently.

Theory-Guided Prediction:
1. Standard inventory theory suggests there is an optimal order quantity balancing overstock and stockout risk.
2. Review the participant's behavior in the first 15 rounds.
3. For rounds 16 to 30, choose order quantities that aim to maximize profit using both theory and participant history.
\end{lstlisting}


\end{document}